\documentclass[12pt,letter,showpacs,amsmath,amssymb]{article}
\usepackage{graphicx}% Include figure files
\usepackage{dcolumn}% Align table columns on decimal point
\usepackage{bm}% bold math
\begin{document}

\title{Born-Oppenheimer Breakdown in Graphene}

\author{Simone Pisana$^1$, Michele Lazzeri$^2$, Cinzia Casiraghi$^1$, Kostya S. Novoselov$^3$,\\Andre K. Geim$^3$, Andrea C. Ferrari$^1$\footnote{acf26@eng.cam.ac.uk}, Francesco Mauri$^2$\footnote{francesco.mauri@impmc.jussieu.fr}} \maketitle
$^1$Engineering Department, Cambridge
University, 9 JJ Thomson Avenue, Cambridge CB3 0FA,UK\\
$^2$IMPMC, Universit\'es Paris 6 et 7, CNRS, IPGP, 140 rue de
Lourmel, 75015 Paris, France\\
$^3$Department of Physics and Astronomy, University of
Manchester, Manchester, M13 9PL, UK

{\bf The Born-Oppenheimer approximation (BO)~\cite{BornO} is the
standard ansatz to describe the interaction between electrons and
nuclei.  BO assumes that the lighter electrons adjust adiabatically
to the motion of the heavier nuclei, remaining at any time in their
instantaneous ground-state. BO is well justified when the energy gap
between ground and excited electronic states is larger than the
energy scale of the nuclear motion.  In metals, the gap is zero and
phenomena beyond BO (such as phonon-mediated superconductivity or
phonon-induced renormalization of the electronic properties)
occur~\cite{grimvall}.  The use of BO to describe lattice motion in
metals is, therefore, questionable~\cite{ponosov,white}. In spite of
this, BO has proven effective for the accurate determination of
chemical reactions~\cite{scheffler}, molecular
dynamics~\cite{alfe,marzari} and phonon
frequencies~\cite{chester,Baroni,savrasov} in a wide range of
metallic systems. Graphene, recently discovered in the free
state~\cite{Novpnas,Novsci04}, is a zero band-gap
semiconductor~\cite{wallace}, which becomes a metal if the Fermi
energy is tuned applying a gate-voltage
$V_g$~\cite{Novo05,Novsci04}. Graphene electrons near the Fermi
energy have two-dimensional massless dispersions, described by Dirac
cones. Here we show that a change in $V_g$ induces a stiffening of
the Raman G peak (i.e. the zone-center E$_{2g}$ optical
phonon~\cite{ferrari06,tuinstra}), which cannot be described within
BO.  Indeed, the E$_{2g}$ vibrations cause rigid oscillations of the
Dirac-cones in the reciprocal space~\cite{Dubay}. If the electrons
followed adiabatically the Dirac-cone oscillations, no change in the
phonon frequency would be observed. Instead, since the
electron-momentum relaxation near the Fermi
level~\cite{kim06,hertel,perfetti} is much slower than the phonon
motion, the electrons do not follow the Dirac-cone displacements.
This invalidates BO and results in the observed phonon stiffening.
This spectacular failure of BO is quite significant since BO has
been the fundamental paradigm to determine crystal vibrations from
the early days of quantum
mechanics~\cite{BornO,chester,pick,Baroni,savrasov}.}

\begin{figure}
\centerline{\includegraphics[width=130mm]{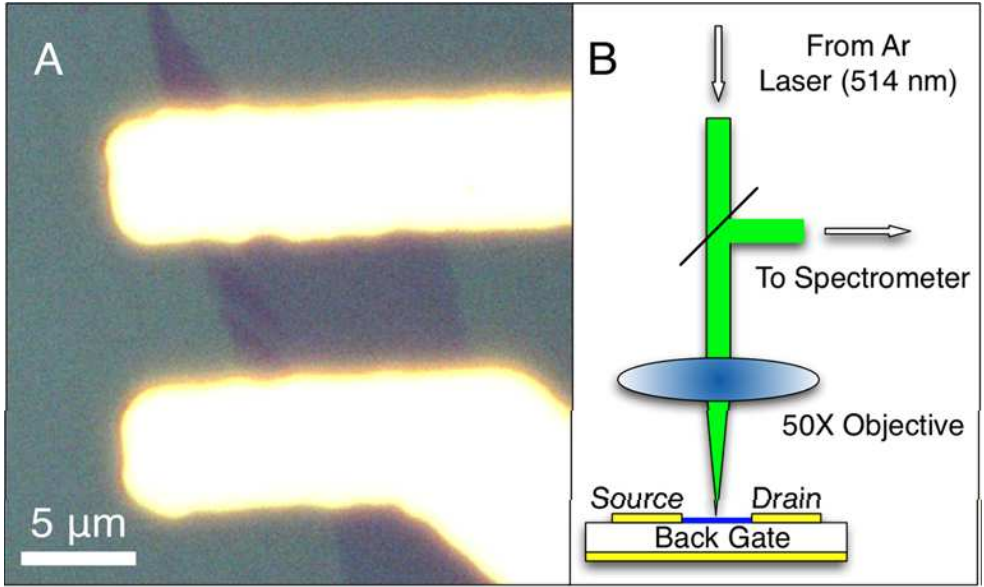}}\caption{(A)
Optical micrograph of the contacted graphene sample. (B) Schematic
of the Raman+transport measurement system} \label{fig1}
\end{figure}

Graphene samples are prepared by micromechanical cleavage of bulk
graphite at the surface of an oxidized Si wafer with a 300 nm thick
oxide layer, following the procedures described in
Ref.~\cite{Novpnas}. This allows us to obtain graphene monocrystals
exceeding 30 microns in size. By using photolithography, we then
make Au/Cr electrical contacts, which enable the application of a
gate voltage, V$_g$, between the Si wafer and graphene
(Fig.~\ref{fig1}A,B). The resulting devices are characterized by
electric-field-effect measurements~\cite{Novsci04, Novo05,Kim},
yielding a charge carrier mobility $\mu$ of 5,000 to 10,000
cm$^2$/Vs at 295K and a zero-bias (V$_g$=0) doping of
$\sim$10$^{12}$ cm$^{-2}$~\cite{schedin}. This is reflected in the
existence of a finite gate voltage V$_n$ at which the Hall
resistance is zero  and the longitudinal resistivity reaches its
maximum. Accordingly, a positive (negative) V$_g$-V$_n$ induces an
electron (hole) doping, having an excess-electron
surface-concentration of n=$\eta$(V$_g$ - V$_n$). The coefficient
$\eta\approx$7.2~10$^{10}$cm$^{-2}/$V is found from Hall effect
measurements and agrees with the geometry of the resulting
capacitor~\cite{Novsci04,Novpnas,Novo05}.

\begin{figure}
\centerline{\includegraphics[width=105mm]{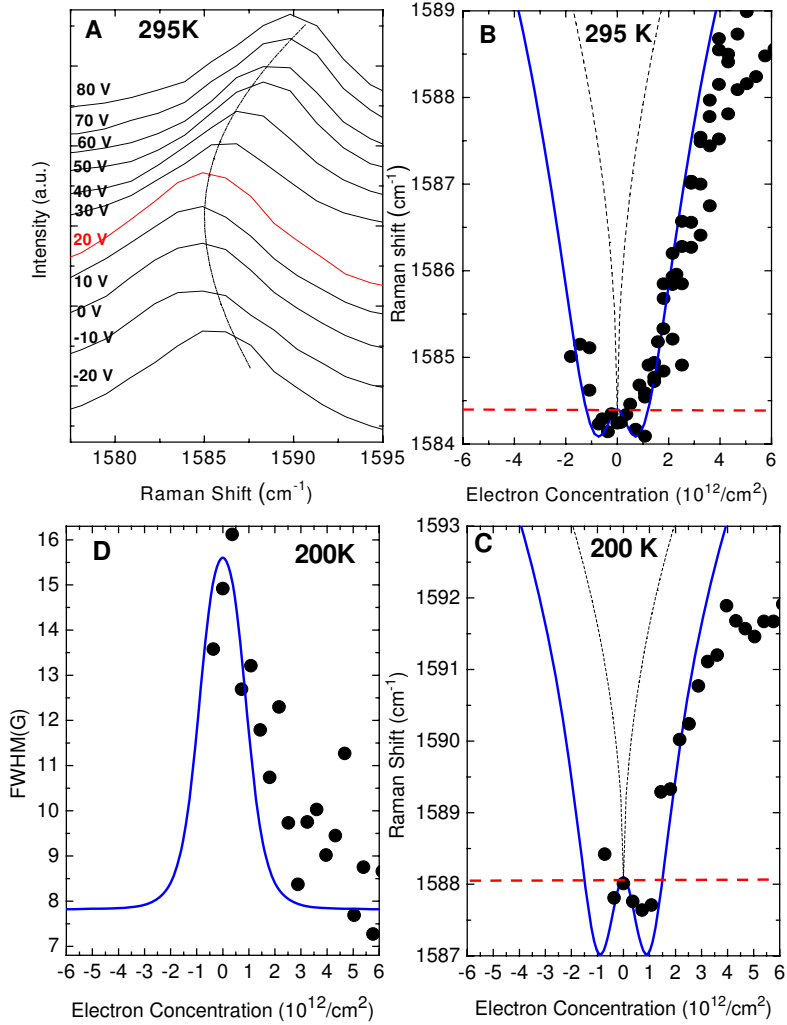}}
\caption{(A) G peak of graphene measured at room temperature as a
function of V$_g$. The red spectrum corresponds to the Dirac
point.(B-C) G peak position as a function of electron concentration
at 200 and 295~K: (black dots) measurements; (red-dashed line)
adiabatic Born-Oppenheimer; (blue line) finite-temperature
non-adiabatic calculation from Eq.~\ref{eq6}; (thin dashed black)
simplified non-adiabatic calculation from Eq.~\ref{eq5}. (D) G peak
Full Width at Half Maximum, FWHM(G), at 200K as a function of
electron concentration: (dots) measured; (Blue line) theoretical
FWHM of a Voigt profile obtained from a Lorentzian component given
by Eq.~\ref{eq7}, and a constant Gaussian component of
$\sim$8~cm$^{-1}$.} \label{fig2}
\end{figure}

Unpolarized Raman spectra are measured at 295 and 200~K in ambient
air and in vacuum ($<$5~10$^{-6}$ mbar), respectively, with a
Renishaw spectrometer at 514nm using a 50$\times$ long working
distance objective, Fig.~\ref{fig1}B. The incident power is kept
well below 4mW in order to avoid sample damage or laser induced
heating~\cite{ferrari06}. The Raman spectra are measured as a
function of the applied V$_g$, Fig.~\ref{fig2}A. Each spectrum is
collected for 30 seconds. The applied gate voltage tends to move the
Dirac point, especially at room temperature. We thus determine the
V$_g$ corresponding to the minimum G peak position, and use this to
estimate V$_n$. Fig.~\ref{fig2}A,B show that the G peak upshifts
with positive applied V$_g$-V$_n$. It also shows a similar trend,
albeit over a smaller voltage range, for negative V$_g$-V$_n$. This
upshift for both electron and hole doping is qualitatively similar
to that reported by Yan {\it et al.} for electrically doped graphene
measured at 10K~\cite{Pinczuk}.

The Raman G peak of graphene corresponds to the E$_{2g}$ phonon at
${\bm \Gamma}$~\cite{ferrari06,tuinstra}. Phonon calculations done
within BO for undoped graphene and graphite, show the presence of a
Kohn anomaly in the phonon dispersion of the E$_{2g}$ mode near
$\bf{\Gamma}$~\cite{piscanec04}. A Kohn anomaly is the softening of
a phonon of wavevector ${\bf q} \sim 2{\bf k}$$_F$, where ${\bf
k}$$_F$ is a Fermi surface wavevector~\cite{kohn}. By doping
graphene, intuitively one could expect that the change in the Fermi
surface should move the Kohn anomaly away from $\textbf{q=0}$ and,
thus, stiffen the $\bf{\Gamma}$ phonon detected by Raman
measurements, which would be in agreement with our experiments. To
validate this picture, we need to compute the frequency of the
E$_{2g}$ mode in doped graphene.

\begin{figure}
\centerline{\includegraphics[width=100mm]{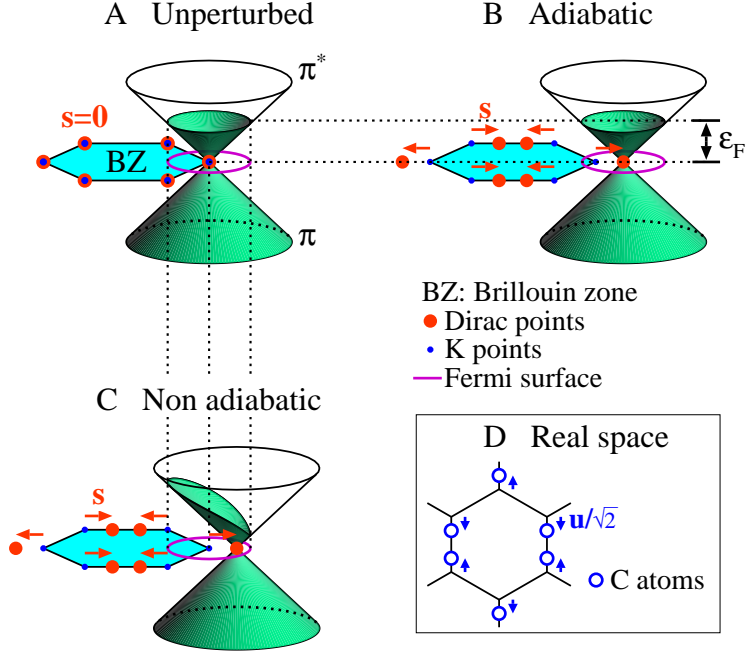}} \caption{
Schematic $\pi$ band structure of doped graphene near the high
symmetry {\bf K} point of the Brillouin zone.  The filled electronic
states are colored in green.  (A) Bands of the perfect crystal. The
Dirac point coincides with {\bf K}, the electronic states are
filled up to the Fermi energy $\epsilon_F$ and the Fermi surface is
a circle centered at {\bf K}. (B) Bands in presence of an E$_{2g}$
lattice distortion.  The Dirac points are displaced from {\bf K} by
$\pm {\bf s}$.  In the adiabatic approximation, the electrons remain
in the instantaneous ground state: the bands are filled up to
$\epsilon_F$ and the Fermi surface follows the Dirac-point
displacement. The total electron-energy does not depend on {\bf s}.
(C) Bands in presence of an E$_{2g}$ lattice distortion.  In the
non-adiabatic case, the electrons do not have time to relax their
momentum (through impurity, electron-electron and electron-phonon
scattering) to follow the instantaneous ground state. In absence of
scattering, the electron momentum is conserved and a state with
momentum {\bf k} is occupied if the state with the same {\bf k} is
occupied in the unperturbed case. As a consequence, the Fermi
surface is the same as in the unperturbed case and does not follow
the Dirac-cone displacement. The total electron-energy increases
with $s^2$ resulting in the observed E$_{2g}$-phonon stiffening. (D)
Atomic pattern of the E$_{2g}$ lattice distortion, corresponding to
the Raman G peak.  The atoms are displaced from the equilibrium
positions by $\pm {\bf u}/\sqrt{2}$. Note that the displacement
pattern of the Dirac points (in reciprocal space) is identical to
the displacement pattern of the carbon atoms (in real space). }
\label{fig3}
\end{figure}

In graphene, the electronic bands near the high-symmetry {\bf K}
points are well described by a Dirac dispersion~\cite{wallace}
$\epsilon({\bf k},\pi^*)=\hbar v_F k$ and $\epsilon({\bf
k},\pi)=-\hbar v_F k$, where ${\bf k+K}$ is the momentum of the
Dirac Fermions, $v_F$ is the Fermi velocity and $\hbar
v_F=5.52$~eV~\AA, from density functional theory (DFT)
calculations~\cite{piscanec04} (Fig.~\ref{fig3}A). The Dirac point
is defined by the crossing of these conic bands and coincides with
${\bf K}$, Fig.~\ref{fig3}A. Thus, at zero temperature, the
doping-induced shift of the Fermi level from the Dirac-point is
$\epsilon_F={\rm sgn}(n) \sqrt{n\pi} \hbar v_F$, where ${\rm
sgn}(x)$ is the sign of $x$.

The E$_{2g}$ phonon in graphene consists of an in-plane displacement
of the carbon atoms by a vector $\pm{\bf u}/\sqrt{2}$ as sketched in
Fig.~\ref{fig3}D. In presence of such atomic displacements, the
bands are still described by a cone (i.e. a gap does not open) with
the Dirac-point shifted from ${\bf K}$ by a vector ${\bf s}$
(Fig.~\ref{fig3}B,C)~\cite{Dubay}. In practice, the atomic-pattern
of the E$_{2g}$ vibrations is mirrored into an identical pattern of
Dirac-point vibrations in the reciprocal space. The dependence of
the electronic-bands on {\bf u} can be obtained (see supplementary information) from the DFT
electron-phonon coupling matrix-elements (Eq.~6 and note 24 of
Ref.~\cite{piscanec04}):
\begin{equation}
\epsilon({\bf k},\pi^*/\pi,{\bf u})=\pm\hbar v_F |{\bf k}-{\bf s}({\bf u})|
\label{eq1}
\end{equation}
where ${\bf s}\cdot{\bf u}=0, s=u\sqrt{2\langle D^2_{\bm \Gamma}
\rangle_F}/(\hbar v_F)$, and  $\langle D^2_{\bm \Gamma}
\rangle_F=45.6$~(eV)$^2$/\AA$^{-2}$ is the deformation potential of
the E$_{2g}$ mode~\cite{lazzeri06}.
Eq.~\ref{eq1} well reproduces the modification of the DFT band structure of
graphene due to a static displacement (frozen-phonon) of the
atoms according to the G phonon pattern.

The knowledge of the electronic-bands (in the presence of a phonon)
allows the determination of the phonon energy
$\hbar\omega_{\epsilon_F}$ as a function of $\epsilon_F$. In particular,
\begin{equation}
\hbar\Delta\omega = \hbar\omega_{\epsilon_F} - \hbar\omega_0 =
\frac{\hbar}{2M\omega_0}
\frac{d^2\Delta E}{(du)^2},
\label{eq2}
\end{equation}
where $M$ is the carbon mass, $\Delta\omega \ll \omega_0$ and
$\Delta E$ is the variation of the electronic energy with $\epsilon_F$.

Within BO, $\Delta E(u)$ is computed assuming a static atomic
displacement. Under this hypothesis, for any given displacement $\bf
u$, the electrons are supposed to be in the ground state
configuration, i.e. the bands $\epsilon({{\bf k},\pi^*,{\bf u}})$
are filled up to $\epsilon_F$ (Fig.~\ref{fig3}B). Thus, the
adiabatic $\Delta E$ is
\begin{equation}
\Delta E(u) = \frac{4 A}{(2\pi^2)}
\int_{\epsilon({{\bf k},\pi^*,{\bf u}})<\epsilon_F}
\epsilon({{\bf k},\pi^*,{\bf u}})~d^2k,
\label{eq3}
\end{equation}
where we consider $\epsilon_F>0$, $A=5.24$~\AA$^2$ is the unit-cell
area, a factor $4$ accounts for spin and ${\bf K}$-point degeneracy.
Combining Eq.~\ref{eq1} and ~\ref{eq3}, we have that $\Delta E$ does
not depend on {\bf u} and $\hbar\Delta \omega=0$. \textit{Thus,
within BO, the Raman G peak position is independent of $\epsilon_F$,
in contrast with the trend reported in Fig.~\ref{fig2}B,C}. Note
that Ref.~\cite{castroneto} calculated the doping dependence of the
G peak position within BO, but found significant phonon softening
for increasing doping, quite the opposite of the experiments, and
in contrast with our BO calculation.

The failure of the frozen-phonon calculation, urges us to re-examine
the assumptions underlying BO. The E$_{2g}$ phonon is a dynamical
perturbation described by a time-dependent lattice displacement
$\tilde{\bf u}(t)={\bf u}\cos(\omega_0 t)$ oscillating at the G peak
frequency. Within BO, it is assumed that, at any given time $t$, the
electrons are in the adiabatic ground state of the instantaneous
band structure $\epsilon({{\bf k},\pi^*,\tilde{\bf u}}(t))$.
However, the period of the G peak vibrations is $\sim$21 fs, which
is much shorter than the typical electron-momentum relaxation times
$\tau_m$ (due to impurity, electron-electron and electron-phonon
scattering). Indeed, a $\tau_m$ of a few hundreds fs is deduced from
the electron-mobility in graphene~\cite{kim06} and from ultra-fast
spectroscopy in graphite~\cite{hertel,perfetti}. As a consequence
{\it the electrons do not have time to relax their momenta to reach
the instantaneous adiabatic ground state, as assumed in BO.} The
departure from the adiabatic ground state can be accounted for in
the calculation of  $\Delta E$, by filling the perturbed bands,
$\epsilon({{\bf k},\pi^*,{\bf u}})$ with the occupations of the
unperturbed bands $\epsilon({{\bf k},\pi^*,{\bf 0}})$, as in
Fig.~\ref{fig3}C:
\begin{equation}
\Delta E(u) = \frac{4 A}{(2\pi^2)}
\int_{\epsilon({{\bf k},\pi^*,{\bf 0}})<\epsilon_F}
\epsilon({{\bf k},\pi^*,{\bf u}})~d^2k +O(u^3).
\label{eq4}
\end{equation}
This equation is valid in the limit $\epsilon_F\gg\hbar\omega_0/2$,
and can be rigorously derived using time dependent perturbation
theory (see supplementary information). In this case, the non-adiabatic energy,
$\Delta E$, depends on $u$. By replacing Eq.~\ref{eq1} and
Eq.~\ref{eq4} in Eq.~\ref{eq2} and performing the integral we get:
\begin{equation}
\hbar\Delta\omega=
\frac{\hbar A \langle D^2_{\bm \Gamma}\rangle_F}
{\pi M \omega_0 (\hbar v_F)^2}|\epsilon_F|=\alpha' |\epsilon_F|,
\label{eq5}
\end{equation}
where $\alpha'=4.39~10^{-3}$.

The result of Eq.~\ref{eq5} can be extended to any $\epsilon_F$ and
finite temperature $T$ using time dependent perturbation
theory~\cite{lazzerimauri}
to obtain:
\begin{equation}
\hbar\Delta\omega = \alpha' {\rm P} \int_{-\infty}^{\infty}
\frac
{[f(\epsilon-\epsilon_F)-f(\epsilon)] \epsilon^2 {\rm sgn}(\epsilon) }
{\epsilon^2 - (\hbar\omega_0)^2/4} d\epsilon,
\label{eq6}
\end{equation}
where ${\rm P}$ is the principal part, and $f$ is the Fermi-Dirac
distribution at $T$. Fig.~\ref{fig2}B,C shows the excellent
agreement of our non-adiabatic finite T calculation (Eq.~\ref{eq6})
with the experiments. The measured trends are also captured by the
simplified model, Eq.~\ref{eq5}. By comparing the predictions of the
BO calculation and of the non-adiabatic model, we conclude that
{\it the stiffening of the E$_{2g}$ mode with $|\epsilon_F|$ is due
to the departure of the electron population from the adiabatic
ground state}.

A pictorial interpretation of this phenomenon can be obtained by
considering what happens to a filled glass when shaken horizontally.
The liquid gravitational-energy and its level mimic the electronic
energy $\Delta E$ and $\epsilon_F$, respectively. The shaking
frequency mimics the phonon frequency and the relaxation time of the
liquid-surface mimics the electron relaxation time. If the motion of
the glass is slow, the liquid surface remains flat and its
gravitational-energy is independent of the glass horizontal
position, as in Eq.~\ref{eq3} and in Fig.~\ref{fig3}B. If the motion
of the glass is rapid, the liquid surface profile is not flat and
its gravitational-energy increases with the displacement of the
glass, as in Eq.~\ref{eq4} and Fig.~\ref{fig3}C. To push the analogy
even further, one should use a non-cylindrical glass, for which the
liquid surface increases with the liquid level. In this case, the
higher the liquid-level, the larger the difference between the
gravitational energies in the fast- and slow-shaken glass. Indeed,
in graphene, the higher the Fermi level, the larger the difference
between the non-adiabatic $\Delta E$ and the adiabatic $\Delta E$.
This causes the observed stiffening of the phonon frequency with
$\epsilon_F$.

The validity of our model is further confirmed by the analysis of
the G peak linewidth. The phonon decaying into an electron-hole pair
gives the most important contribution to the homogeneous broadening
of the E$_{2g}$ phonon. The full-width at half-maximum, $\gamma$,
can be computed extending to finite $T$ and $\epsilon_F\ne0$ the
results of Ref.~\cite{lazzeri06}:
\begin{equation}
\gamma =
%\frac{\pi\hbar\omega_0\alpha'}{2} <- con questo hai energia
%\frac{\pi^2\omega_0\alpha'}{c} <- con questo hai cm-1
\frac{\pi^2\omega_0\alpha'}{c}
\left[f\left( -\frac{\hbar\omega_0}{2}-\epsilon_F\right)
-f\left( \frac{\hbar\omega_0}{2}-\epsilon_F\right)\right],
\label{eq7}
\end{equation}
where $c$ is the speed of light. At $T=0$, $\gamma=11$~cm$^{-1}$ for
$\epsilon_F=0$ and $\gamma$ drops to zero for
$\epsilon_F>\hbar\omega_0/2$ because the scattering process is
forbidden by the Pauli exclusion principle~\cite{lazzeri06}.
Fig.~\ref{fig2}D shows a good agreement between the experimental and
theoretical $\gamma$, once a constant inhomogeneous Gaussian
broadening of $\sim$~8~cm$^{-1}$ is added to the electron-phonon
contribution of Eq.~\ref{eq7}.

Concluding, the observed stiffening of the E$_{2g}$ phonon in doped
graphene represent a spectacular failure of the adiabatic
Born-Oppenheimer approximation. Within BO, the energy of a
zone-center phonon is determined by two contributions: the
distortion of the electronic bands, associated with the phonon
displacement, and  the consequent rearrangement of the Fermi
surface. These two contributions cancel out exactly in graphene
because of the peculiar rigid motion of the Dirac-cones, associated
to the E$_{2g}$ phonon. In general, a correct phonon treatment
\textit{should not} include the BO Fermi-surface rearrangement,
whenever the electron-momentum relaxation time (near $\epsilon_F$)
is longer than the phonon period. We anticipate the failure of BO,
shown here, to have important consequences in the description of
vibrational properties of carbon-nanotubes and in phonon-mediated
superconductors.

\paragraph{Acknowledgments}

The authors thank P. Kim and A. Pinczuk for useful discussions and
for sending a preprint of Ref.~\cite{Pinczuk}. A. C. F. acknowledges
funding from the Royal Society and The Leverhulme Trust.
Calculations were performed at the IDRIS supercomputing center.

\newpage

\appendix{\Huge \bf Supplementary Information}
\section*{Derivation of Eq.~\ref{eq1}}

The electronic Hamiltonian for the $\pi$,$\pi^*$ basis can be
written as a $2\times2$ matrix:
\begin{equation}
H({\bf k,0}) =
\left( \begin{array}{cc}
\hbar v_F k & 0 \\
0 & -\hbar v_F k \end{array}\right),
\label{aeq1}
\end{equation}
where {\bf k} is a small in plane wave-vector and ${\bf K+k}$ is the
electronic momentum. Let us consider a distortion of the lattice
according to a ${\bm \Gamma}-E_{2g}$ phonon pattern (note that the
${\bm \Gamma}-E_{2g}$ phonon is doubly degenerate). At the lowest
order the $\pi$-bands Hamiltonian changes as
\begin{equation}
H({\bf k,u}) = H({\bf k,0}) +
\frac{\partial H({\bf k,0})}{\partial u}
u
\label{aeq2}
\end{equation}
where $u$ is the phonon normal coordinate (the two atoms in the
unit-cell are displaced by $\pm u/\sqrt2$ along a given direction in
the plane). $\partial H /(\partial u)$ can be obtained from the
ab-initio deformation potential matrix elements. Following Ref.~\cite{piscanec04} 
(Eq.6 and note 24) and Ref.~\cite{lazzeri06}, for the E$_{2g}$ phonon mode and for
a small {\bf k}
\begin{eqnarray}
\left|\langle {\bf k}\pi^*|\frac{\partial H}{\partial u}|{\bf k}\pi^*\rangle
\right|^2
&=&
\left|\langle {\bf k}\pi|\frac{\partial H}{\partial u}|{\bf k}\pi\rangle
\right|^2
=\langle D^2_{\bm \Gamma}\rangle_F
[1+\cos(2\theta)] \label{aeq3}\\
\left|\langle {\bf k}\pi^*|\frac{\partial H}{\partial u}|{\bf k}\pi\rangle
\right|^2
&=&
\langle D^2_{\bm \Gamma}\rangle_F
[1-\cos(2\theta)], \label{aeq4}
\end{eqnarray}
where $|{\bf k}\pi/\pi^* \rangle$ are the electronic states with momentum
{\bf K+k} and $\theta$ is the angle between {\bf k} and the direction
perpendicular to the atomic vibration.
Taking the square root of Eqs.~\ref{aeq3},~\ref{aeq4} and inserting them into
Eq.~\ref{aeq2}
\begin{equation}
H({\bf k,u}) =
\left( \begin{array}{cc}
\hbar v_F k & 0 \\
0 & -\hbar v_F k 
\end{array} \right)
+
\sqrt{2\langle D^2_{\bm \Gamma}\rangle_F}
\left( \begin{array}{cc}
\cos (\theta) & \sin(\theta) \\
\sin(\theta) & -\cos (\theta) 
\end{array} \right)
u.
\label{aeq5}
\end{equation}
The eigenvalues of Eq.~\ref{aeq5} are then
\begin{equation}
\epsilon=\pm\hbar v_F
\sqrt{ k^2 + s^2 + 2ks\cos(\theta)}
=\pm \hbar v_F |{\bf k}-{\bf s}({\bf u})|,
\end{equation}
where {\bf s} is defined in the main text.

\section*{Derivation of Eq.~\ref{eq4}}

Considering the Taylor expansion of $\Delta E(u)$ in $u$,
Eq.~\ref{eq4} is equivalent to:
\begin{equation}
\frac{d^2}{(du)^2}\Delta E(u)=
\frac{d^2}{(du)^2}
\left\{
\frac{4A}{(2\pi)^2}
\int_{\epsilon({\bf k},\pi^*,{\bf 0})<\epsilon_F}
\epsilon({\bf k},\pi^*,{\bf u}) ~d^2k
\right\}.
\end{equation}
In this section, we will demonstrate that
\begin{equation}
\hbar\Delta\omega = \frac{\hbar}{2M\omega_0}
\frac{d^2}{(du)^2}
\left\{
\frac{4A}{(2\pi)^2}
\int_{\epsilon({\bf k},\pi^*,{\bf 0})<\epsilon_F}
\epsilon({\bf k},\pi^*,{\bf u}) ~d^2k
\right\},
\label{eqb1}
\end{equation}
at $T=0$ and under the condition
\begin{equation}
\epsilon_F\gg \hbar\omega_0/2.
\label{cond}
\end{equation}
Using Eq.~\ref{eq2}, Eq.~\ref{eq4} will then immediately follow.

Within time dependent perturbation theory, $\hbar\Delta\omega$ is
(see Eq. 10 of Ref.~\cite{lazzerimauri}):
\begin{equation}
\hbar\Delta\omega=\frac{\hbar}{2M\omega_0}[F^{\epsilon_F}_{\bf 0}(\omega_0)-
F^{0}_{\bf 0}(\omega_0)].
\label{diff}
\end{equation}
were at $T=0$,
\begin{equation}
F^{\epsilon_F}_{\bf 0}(\omega_0)=
\frac{2}{N_{\bf k}}\sum_{{\bf k},o,e}
|D_{{\bf k}o,{\bf k}e}|^2
\left\{
\frac{1}{\epsilon_{{\bf k}o}-\epsilon_{{\bf k}e}+\hbar\omega_0}
+\frac{1}{\epsilon_{{\bf k}o}-\epsilon_{{\bf k}e}-\hbar\omega_0}
\right\}.
\label{eqb2}
\end{equation}
Here the index $o$ and $e$ denotes the occupied ($\epsilon_{{\bf k}o}< \epsilon_F$) and empty bands ($\epsilon_{{\bf k}e}> \epsilon_F$), and
\begin{equation}
D_{{\bf k}o,{\bf k}e}=\langle {\bf k}o|\frac{\partial H}{\partial u}|{\bf k}e\rangle.
\end{equation}
Now we consider only the $\pi$ and $\pi^*$ bands and we substitute
$1/N_{\bf k}\sum_{\bf k}$ with $A/(2\pi)^2\int d^2k$, where $A$ is
the unit-cell area and the integral is restricted on a circle of
radius $\bar{k}$, centered on {\bf K}. Assuming a Dirac dispersion
for the $\pi$ and $\pi^*$ bands, $\epsilon_{{\bf k}e}-\epsilon_{{\bf
k}o}\ge 2\epsilon_F$. Thus, if the condition of Eq.~\ref{cond}
holds, $|\epsilon_{{\bf k}e}-\epsilon_{{\bf k}o}|\gg \hbar\omega_0$
and the $\hbar\omega_0$ in the denominators of Eq.~\ref{eqb2} can be
neglected. Eq.~\ref{eqb2} becomes
\begin{equation}
F^{\epsilon_F}_{\bf 0} = \frac{8A}{(2\pi)^2}
\int_{\epsilon({\bf k},\pi^*,{\bf 0})>\epsilon_F,k<{\bar k}}
\frac{|D_{{\bf k}\pi^*,{\bf k}\pi}|^2}{
\epsilon({\bf k},\pi,{\bf 0})-
\epsilon({\bf k},\pi^*,{\bf 0})}
~d^2k,
\end{equation}
where $\epsilon({\bf k},\pi/\pi^*,{\bf 0})$ are the bands of the undistorted
graphene structure.
From Eq.~\ref{diff},
\begin{equation}
\hbar\Delta\omega=
\frac{\hbar}{2M\omega_0}
\left\{
-\frac{8A}{(2\pi)^2}
\int_{\epsilon({\bf k},\pi^*,{\bf 0})<\epsilon_F}
\frac{|D_{{\bf k}\pi^*,{\bf k}\pi}|^2}{
\epsilon({\bf k},\pi,{\bf 0})-
\epsilon({\bf k},\pi^*,{\bf 0})}
~d^2k
\right\}.
\label{eqb4}
\end{equation}
From textbook static second order perturbation theory
\begin{equation}
\frac{1}{2}
\left. \frac{d^2\epsilon({\bf k},\pi^*,{\bf u})}{(du)^2}
\right|_{u=0}
=\frac{|D_{{\bf k}\pi^*,{\bf k}\pi}|^2}{
\epsilon({\bf k},\pi^*,{\bf 0})-
\epsilon({\bf k},\pi,{\bf 0})}.
\label{eqb5}
\end{equation}
Substituting Eq.~\ref{eqb5} in Eq.~\ref{eqb4} we have
\begin{equation}
\hbar\Delta\omega=
\frac{\hbar}{2M\omega_0}
\left\{
\frac{4A}{(2\pi)^2}
\int_{\epsilon({\bf k},\pi^*,{\bf 0})<\epsilon_F}
\frac{d^2\epsilon({\bf k},\pi^*,{\bf u})}{(du)^2}
~d^2k
\right\}.
\label{eqb6}
\end{equation}
Eq.~\ref{eqb1} is, finally, obtained by taking the derivation with
respect to $u$ in Eq.~\ref{eqb6} outside the integral.

\end{document}